\documentclass[aps,prl,twocolumn,superscriptaddress]{revtex4-1}

\newcommand{\ket}[1]{ | #1 \rangle }
\newcommand{\bra}[1]{ \langle #1 | }
\newcommand{\braket}[2]{\left \langle #1 \middle| #2 \right \rangle}
\newcommand{\braketmatrix}[3]{\left \langle #1 \middle| #2 \middle| #3 \right \rangle}
\usepackage{amsmath}
\usepackage{graphicx}
\begin{document}

\title{Powering a quantum clock with a non-equilibrium steady state}

\author{Daniele Nello}
\affiliation{Sissa, Via Bonomea, 265 Trieste (Italy)}
\author{Alessandro Silva}
\affiliation{Sissa, Via Bonomea, 265 Trieste (Italy)}
\email{danello@sissa.it}

\begin{abstract}
   We propose powering a quantum clock with the non-thermal resources offered by the stationary state of an integrable quantum spin chain, driven out of equilibrium by a quench in a parameter of our choice. Analyzing the bias conditions of the clock, we establish a direct connection with the negativity of the steady-state response function. Using experimentally relevant examples of quantum spin chains, we suggest crossing a phase transition point is crucial for optimal performance. The coupling takes place through a global observable and, even in this case, the battery lifespan is found to be extensive in its size.

\end{abstract}

\maketitle
Quantum clocks are timekeeping devices that exploit quantum phenomena to measure time intervals with extraordinary accuracy. They are realized in a variety of nanoscale devices and platforms, ranging from quantum dots to lasers and thermal machines \cite{milburn}. The performance of these clocks, however, comes at a thermodynamic cost: higher accuracy necessitates greater entropy production, reflecting fundamental physical limits \cite{erker,huber}.

In general, all clocks are non-equilibrium dissipative systems, relying on an external free-energy source \cite{milburn,timeandclocks,gangat,he2023effect,manikandan2023autonomous}. A quantum clock can be operated using the smallest quantum thermal machine, which was introduced by Brunner et.al. \cite{popescu}. It consists of a virtual qubit, obtained from a subspace of the combined Hilbert space of two qubits, each in thermal equilibrium with a different thermal bath at a fixed temperature. 
A population inversion is created in the virtual qubit when its temperature becomes negative, an effect that can be exploited to perform work on a register consisting of a ladder of a finite number $d$ of energy levels~\cite{erker}. Each time the ladder reaches the top, a photon is spontaneously emitted, thus constituting the clock signal. 

In this Letter, we suggest that a quantum clock can be designed and operated by exploiting the non-passivity of the non-thermal stationary state of an integrable quantum spin chain, driven out from equilibrium by a quench in one of its parameters. This stationary state, drawn from a generalized Gibbs ensemble (GGE) \cite{calabrese2,foini,vidmar}, in general has a finite ergotropy \cite{alicki,allahverdian} and is consequently a particular case of a quantum battery~\cite{alicki,campaioli2018quantum,rossini,piccitto}. While the latter is typically characterized by investigating charging power and quantifying extractable work~\cite{campaioli2018quantum,quach2023quantum,Barra,catalano} here we study in detail the conditions upon which work can be used to operate a specific quantum clock. In particular, coupling the battery to a qubit, which in turn drives transitions in a \it d\rm-level ladder, 
we examine the conditions in the parameter space guaranteeing the clock operation and establish a direct link with the negativity of the imaginary part of the steady-state response function, a quantity linked to energy extraction from such a system \cite{rossini,piccitto}.
We also examine experimentally relevant examples, that can be simulated using trapped ions \cite{Kim,rossini,qiao}. In these examples, we find the requirement of the crossing of the critical point for the clock's bias condition, bearing a striking connection with the optimal charging protocol of Ref. \cite{grazi}.



In the present work, a spin chain taken out of equilibrium will act as the quantum battery. Hence, overall the Hamiltonian describing all the components comprises five distinct terms
\begin{equation}\label{Hamiltonian}
\hat{H} = \hat{H}_B + \hat{H}_Q + \hat{H}_{QB} + \hat{H}_L + \hat{H}_{QL} .
\end{equation}
Starting from the left, we have the Hamiltonian of the aforementioned battery $\hat{H}_B$, an integrable spin chain.  The battery is charged at time $t=0$ by quenching one of its parameters from $\lambda_i$ to $\lambda_f$. Subsequently, after it reaches a stationary state, we couple to it a qubit with spacing $\epsilon_0$
\begin{equation}
\hat{H}_Q = \frac{\epsilon_0}{2} \hat{\sigma}_Q^z,
\end{equation}
through a specific observable of the battery, whose choice is crucial for the clock's operation. In terms of a generic observable $\hat{A}$, the qubit-battery interaction term has the expression
\begin{equation}
    \hat{H}_{QB} = \hat{\sigma}_Q^x \hat{A}.
\end{equation}

The ladder of d energy levels $\hat{H}_L$ reads
\begin{equation}
\hat{H}_L = \sum_{k=0}^{d-1} k\epsilon_w \hat{c}^\dag_k \hat{c}_k,
\end{equation}
with the resonance condition $\epsilon_w = \epsilon_0$, ensuring that the levels $\ket{1}_Q \ket{k}_L$ have the same energy as $\ket{0}_Q \ket{k+1}_L$ \cite{popescu}.
The interaction term between the qubit and the ladder reads
\begin{equation}
\hat{H}_{QL} = g \sum_{k=0}^{d-1} \left( \hat{\sigma}_Q^- \hat{c}^\dag_{k+1} \hat{c}_k + h.c. \right)
\end{equation}
where $g$ represents the coupling strength.

The battery generates a population inversion in the qubit by ensuring the population $p_1$ in the upper level of the qubit $\ket{1}_Q$  is greater than the occupation of the lower energy level $p_0$. In turn, the decay from upper to lower level triggers a transition 
$\ket{k}_L\rightarrow \ket{k+1}_L$. 
By analyzing the stationary state of the qubit dynamics, the condition $p_1>p_0$ requires that
the transition rates $\gamma_{\uparrow}$ ($\gamma_{\downarrow}$) from lower to the upper level (and viceversa) are $\gamma_{\uparrow}>\gamma_{\downarrow}$.
(see Supplementary Material).
In turn, the expression of the qubit transition rates \cite{breuer} relative to the choice of a particular coupling observable $\hat{A}$ is
\begin{equation}\label{qubitrates}
\gamma_{\uparrow,\downarrow} = \frac{1}{4}\int ds e^{\mp i \epsilon_0 s} \text{Tr}[\hat{\rho}_{st}\hat{A}(s)\hat{A}(0)],
\end{equation}
where the stationary state of the battery is given by the diagonal ensemble $\hat{\rho}_{st} = \sum_{k,i} p_k^i \ket{k,i} \bra{k,i}$, with $p_k^i = |\braket{k,i}{\psi_0}|^2$ and $\ket{\psi_0}$ being the initial state. Here, $k$ runs over the energy levels and $i$ is a degeneracy index.

As seen clearly in the definition of $\gamma_{\uparrow,\downarrow}$ (Eq.~(\ref{qubitrates})), when operating a clock we are interested in extracting energy at a specific frequency $\epsilon_0$. Let us therefore distinguish between states \it active \rm at resonance ($\gamma_\uparrow > \gamma_\downarrow$) and states \it passive \rm at resonance ($\gamma_\uparrow \leq \gamma_\downarrow$). 
Using the Lehman representation of Eq.~(\ref{qubitrates}), it is not hard to see that a state is active at resonance when 
\begin{equation}\label{gencondition}
\sum_{l,m , \text{s.t.} , E_l > E_m} (p_l - p_m) \big|\braketmatrix{l}{\hat{A}}{m}\big|^2 \delta(\epsilon_0 - E_l + E_m) > 0.
\end{equation}
Therefore being \it active \rm is equivalent to having $p_l > p_m$ at $\epsilon_0 = E_l - E_m$ for all $\ket{l}, \ket{m}$ such that $E_l > E_m$ and $\big|\braketmatrix{l}{\hat{A}}{m}\big|^2 \neq 0$. In turn, when $p_l \leq p_m$ and all other conditions are the same the state turns \it passive\rm. 

It is interesting to notice that the condition of being \it active \rm at resonance is linked to the negativity of the imaginary part of the response function, a quantity that characterizes the possibility of extracting energy from the battery by coupling to it at the specific energy of the qubit~\cite{rossini}. This happens because the Lehman representation of the imaginary part of the response function of $\hat{A}$ is 
\begin{equation}
\bar{\chi}''(\omega) = \sum_{k,k'} (p_k - p_{k'}) \big|\braketmatrix{k}{\hat{A}}{k'}\big|^2 \delta(\omega + E_k - E_{k'}),
\end{equation}
and clearly 
\begin{equation}\gamma_\uparrow - \gamma_\downarrow = -\bar{\chi}''(\epsilon_0),
\end{equation}

\begin{figure}
    \centering
    \includegraphics[scale=0.5]{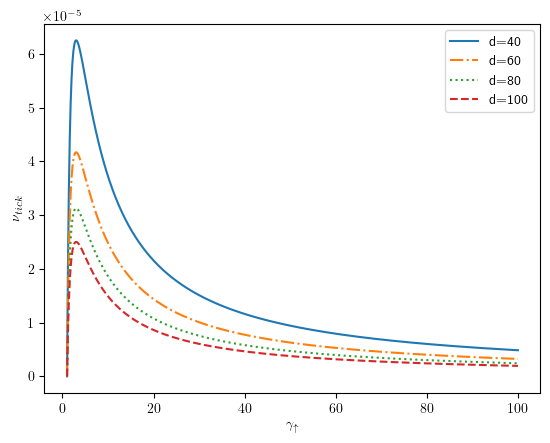}
    \caption{The clock's precision for different values of $d$, being inversely proportional to the ladder dimension.}
    \label{precision}
\end{figure}

\begin{figure}
    \centering
    \includegraphics[scale=0.5]{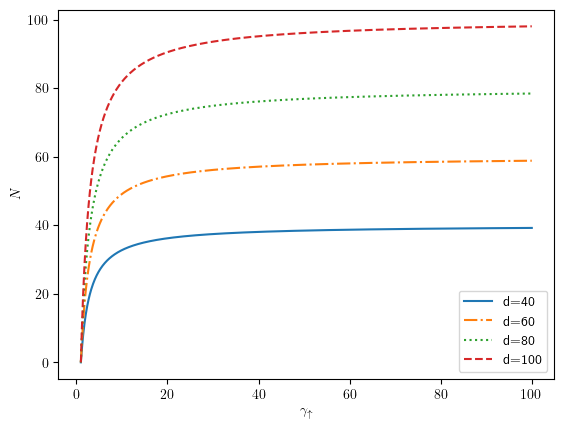}
    \caption{The clock accuracy saturates to a constant value for large bias and is inversely proportional to the ladder's dimension. }
    \label{accuracy}
\end{figure}

Assuming that the battery state is active at resonance, the dynamics of the ladder now unravels similarly to Ref. \cite{erker}. The clock is treated in the weak coupling limit and the ladder dynamics is incoherent. We therefore use to describe it a biased random walk approximation (see Supplementary Material) and calculate the upward and downward transition rates of the ladder as
\begin{equation}
p_\uparrow = 2g^2\frac{\gamma_\uparrow}{(\gamma_\uparrow+\gamma_\downarrow)^2}
\end{equation}
and
\begin{equation}
p_\downarrow = 2g^2 \frac{\gamma_\downarrow}{(\gamma_\uparrow+\gamma_\downarrow)^2} .
\end{equation}
 These results are valid for $g \ll \gamma_+, \Gamma$ , where $\gamma_+=\gamma_\uparrow+\gamma_\downarrow$.
One can see that the ladder's bias $p_\uparrow>p_\downarrow$ implies the condition $\gamma_\uparrow>\gamma_\downarrow$.
This allows us to calculate the accuracy and the precision of the clock, the key metrics for qualifying its performance (Figg.\ref{precision} and \ref{accuracy}). They read
\begin{figure}
    \centering
    \includegraphics[scale=0.5]{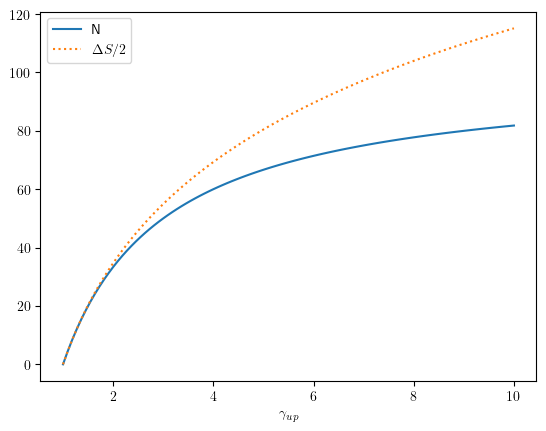}
    \caption{The accuracy of the clock and the entropy production per tick. We see that the two are proportional in the weak bias limit, but in the opposite limit the first saturates, while the second grows logarithmically.}
    \label{entropyandaccuracy}
\end{figure}
\begin{equation}
\nu_{\text{tick}} = \frac{p_\uparrow - p_\downarrow}{d} = \frac{2}{d} \frac{g^2 (\gamma_\uparrow-\gamma_\downarrow)}{(\gamma_\uparrow+\gamma_\downarrow)^2}
\end{equation}
and
\begin{equation}
N = d \left( \frac{p_\uparrow - p_\downarrow}{p_\uparrow + p_\downarrow} \right) = d \frac{\gamma_\uparrow-\gamma_\downarrow}{\gamma_\uparrow+\gamma_\downarrow}.
\end{equation}

Using the results from Ref. \cite{kirchberg}, we compute the entropy production per tick
\begin{equation}
\Delta S_{\text{tick}} = d \ln{\frac{p_\uparrow}{p_\downarrow}} = k_B d \left[ \ln \frac{\gamma_\uparrow}{\gamma_\downarrow} \right].
\end{equation}
This formula reproduces the correct result even in the thermal clock case \cite{erker}. Notably, for $\gamma_\uparrow \simeq \gamma_\downarrow$, we recover the direct relation between entropy production and accuracy (Fig. \ref{entropyandaccuracy}) \cite{erker}
\begin{equation}
N = \frac{\Delta S_{\text{tick}}}{2}
\end{equation}
saturating the TUR bound. The generic relation between these two quantities reads
\begin{equation}
    N =d\tanh\bigg[\frac{\Delta S_{\text{tick}}}{2d}\bigg]
\end{equation}
unchanged from the thermal clock's results. That is to say, being an incoherent quantum clock, the relation derived in Ref. \cite{barato} in the context of Brownian clocks, holds.

Let us now study the practical application of this setup by choosing specific examples for the battery and the coupling.
The first one is the quantum Ising chain in a transverse field. 
Its Hamiltonian is given by
\begin{equation}
    \hat{H}_B=-\sum_{j=1}^L(J^x \hat{\sigma}_j^x \hat{\sigma}_{j+1}^x+J^y \hat{\sigma}_j^y \hat{\sigma}_{j+1}^y)-h\sum_{j=1}^L \hat{\sigma}^z_j.
\end{equation}
Setting $J=1$, the nearest-neighbor coupling can be parameterized as $J_x=\frac{1+\kappa}{2}$ and $J_y=\frac{1-\kappa}{2}$.
At equilibrium, this system
exhibits a quantum phase transition at $h = 1$ from a ferromagnetic to a paramagnetic phase. It is well known that
this model can be diagonalized by mapping the Hamiltonian
onto a system of spinless fermions, and performing in k space a Bogoliubov rotation to end up with a system of free fermions $\hat{\gamma}_k$ with dispersion $\epsilon_k=2\sqrt{(h-\cos k)^2+\kappa^2 \sin^2 k}$ \cite{glen}. Quench dynamics in this model have already been studied in several works, most notably \cite{calabrese1, calabrese2,foini,piccitto}.

\begin{figure}
    \centering
    \includegraphics[scale=0.5]{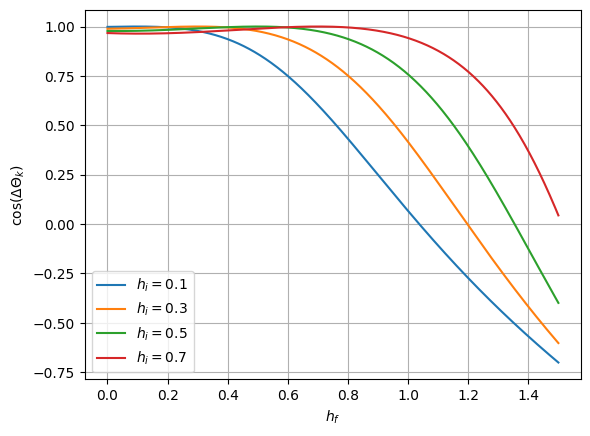}
    \caption{The l.h.s. of Eq. \ref{gencondition} is plotted for different values of $h_i$ in the ferromagnetic phase. The quantity is negative only for $h_f$ in the paramagnetic phase. The solution $k_-$ exists in this range of parameters.}
    \label{condition}
\end{figure}

We choose the coupling between the battery and the qubit as
\begin{equation}
    \hat{H}_{QB}=g_\sigma \hat{\sigma}_Q^x \sum_{j=1}^L \hat{\sigma}_j^z
\end{equation}
and we study quenches in the parameter $h: h_i\rightarrow h_f$ from the ferromagnetic phase. The transition rates are calculated as (see Supplementary Material)
\begin{equation}\label{uptransition}
    \gamma_\uparrow=\frac{2g_\sigma^2}{L\pi}\int_{-\pi/2}^{\pi/2}dk \delta(2|\epsilon_k|-\epsilon_0)\sin^2(2\Theta_{k,f})\sin^2(\Delta\Theta_k)
\end{equation}
and
\begin{equation}\label{downtransition}
    \gamma_\downarrow=\frac{2g_\sigma^2}{L\pi}\int_{-\pi/2}^{\pi/2}dk \delta(2|\epsilon_k|-\epsilon_0)\sin^2(2\Theta_{k,f})\cos^2(\Delta\Theta_k),
\end{equation}
in terms of the difference of Bogoliubov angles $\Delta\Theta_k=\Theta_{k,f}-\Theta_{k,i}$ between the final and initial parameters of the quench. By comparing the two expressions Eq.(\ref{uptransition})-(\ref{downtransition}), we notice the operating condition $\gamma_\uparrow>\gamma_\downarrow$ reads
\begin{equation}
    \cos(2\Delta\Theta_k)|_{k=k^*}<0,
\end{equation}
in terms of the solution $\pm k^*$  of the resonance condition $2|\epsilon_k|=\epsilon_0$ (see Supplementary Material). Parametrizing $\cos k^*= u$, the condition reads
    \begin{equation}\label{cosdeltatheta}
    8\frac{(h_f-u)(h_i-u)+\kappa^2 (1-u^2)}{\epsilon_0\sqrt{\epsilon_0^2/4-4(h_f-u)^2+4(h_i-u)^2}}<0.
\end{equation}
Of course, two additional conditions are provided by the existence of solutions, requiring that the qubit frequency lies within the gap between the two sectors and the maximum range of eigenvalues
\begin{equation}
    \epsilon_0^2/16 \geq (h_f-1)^2
\end{equation}
and 
\begin{equation}
    \epsilon_0^2/16 \leq h_f^2+\kappa^2 .
\end{equation}
Eq. (\ref{cosdeltatheta}) implies that if $h_i<1$, then $h_f>1$, indicating that the quench protocol must cross the phase transition point (Fig. \ref{condition}), as in the optimal charging protocols \cite{grazi}. 
As noted in Ref. \cite{piccitto}, the imaginary part of the response function at the correspondent frequency is negative. The zero of this quantity marks a dynamical quantum phase transition \cite{heyl}, characterized by a population inversion, where the density of quasi-particles above the spectral gap
\begin{equation}
    \langle \hat{n}_k \rangle=\frac{1-\cos(2\Delta\Theta_k)}{2}
\end{equation}
becomes non-thermal ($\langle \hat{n}_k \rangle>\frac{1}{2}$).

A further example of an integrable battery is a model of hard-core bosons hopping in a one-dimensional ring with $L$ sites, which can be mapped onto the XX spin chain \cite{rossini}. Its Hamiltonian is given by
\begin{equation}
    \hat{H}_B(\phi)=t\sum_{j=1}^{L} (e^{i\phi}\hat{b}^\dag_j \hat{b}_{j+1}+h.c.)+V\sum_{j=1}^L (-1)^j \hat{n}_j 
\end{equation}
upon defining $\hat{n}_j=\hat{b}^\dag_j \hat{b}_j$. The creation/annihilation operators are $\hat{b}^\dag_j/\hat{b}_j$ and the field $\phi$ represents an external flux.
The present Hamiltonian has the following spectrum of energy eigenstates
\begin{equation}
    \epsilon_k(\phi)=\sqrt{[2t\cos(k-\phi)]^2+V^2}.
\end{equation}
At $V=0$ there is a critical point, marked by the gap between the two sectors opening. The model thus has a phase diagram comprising an insulating phase at $V\neq 0$ and a superfluid phase at $V=0$.\\

The Hamiltonian for the qubit-battery interaction is
\begin{equation}
    \hat{H}_{QB}=g_J\hat{\sigma}_Q^x \hat{J}_{\phi=0}
\end{equation}
where $g_J$ is the coupling strength and the current in the ring is defined as
\begin{equation}
\hat{J}_{\phi=0} = -\partial_\phi \hat{H}\big|_{\phi=0},
\end{equation}
having the following expression:
\begin{equation}\label{current}
\hat{J}_{\phi=0} = -it \sum_{j=1}^L \left( \hat{b}^\dag_j \hat{b}_{j+1} - \text{h.c.} \right).
\end{equation}
In the Supplementary Material, we calculate the transition rates given by expressions analogous to Eq. \ref{uptransition} and \ref{downtransition}.
The zero solutions of the argument of the delta function are $\pm k^*$, where
\begin{equation}
    k^*=\arccos\sqrt{\frac{\epsilon_0^2/4-V_f^2}{4t^2}}.
\end{equation}
To ensure the existence of a solution to the above relation, the following conditions must be satisfied 
\begin{equation}\label{conditions}
    0<\frac{\epsilon_0^2/4-V_f^2}{4t^2}<1
\end{equation}
This condition requires that $\epsilon_0 $ lies between the gap and the distance between the lowest and highest eigenstates in energy.
\\

\begin{figure}
    \centering
    \includegraphics[scale=0.5]{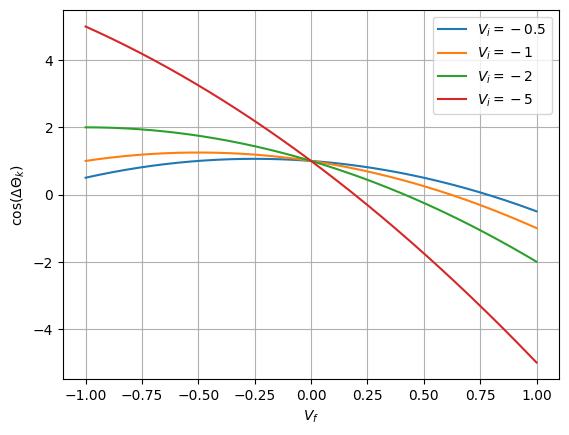}
    \caption{The left-hand side of Eq. \ref{conditione0} plotted for different values of $V_f$. It can be negative only for quenches that cross the critical point $V=0$. The solution 
$k^*$ is acceptable within the considered range of parameters.}
    \label{cosdeltaxx}
\end{figure}

To ensure the proper functioning of the quantum clock, as before, we must require $\gamma_\uparrow>\gamma_\downarrow$. This translates into the condition
\begin{equation}
\cos(2\Delta\Theta_k)|_{k=k^*} < 0,
\end{equation}
i.e.
\begin{equation}\label{conditione0}
\epsilon_0^2/4 - V_f^2 + V_i V_f < 0.
\end{equation}
In terms of the difference of the Bogoliubov angles $\Delta \theta_k=\theta_f-\theta_i$, as defined in the Supplementary Materials, this implies
$V_i V_f<0$. Thus, we recover the key result that the quench must cross the critical point $ V=0$ (see Fig. \ref{cosdeltaxx}).

An important physical question one could ask is how long the resources offered by the battery last. We study the coupling through global observables, as in the previous examples. This battery's lifetime can be determined by considering the average energy dissipation rate, related to the imaginary part of the response function $-\bar{\chi}''(\epsilon_0)$ \cite{Russomanno2013LinearRA}, and the total available energy at the resonance frequency. The latter is extensive in the battery size $L$, as the density of states per unit length is constant.
A rough estimate of this time (see Supplementary Material) gives
\begin{equation}
T^* = -\frac{\gamma_\uparrow+\gamma_\downarrow}{\bar{\chi}''(\epsilon_0)} \frac{L}{2\pi} \rho(\epsilon_0/2) \cos^2(\Delta\Theta_k) \bigg|_{k=k^*},
\end{equation}
where $\rho(\epsilon) = \left| \frac{\partial \epsilon_k}{\partial k} \right|^{-1}$ is the density of states of the spin chain. The most important take-home message is the extensivity of this quantity concerning the battery size. This is good news indeed for the experimental feasibility of our proposal.

In conclusion, in this Letter, we have introduced a minimal model of a quantum clock powered by a quantum battery, specifically an integrable quantum spin chain. Inspired by the thermal clock in Ref.\onlinecite{erker}, we create a non-thermal clock exploiting the resources available from coupling to the stationary state of an integrable system, described by the generalized Gibbs ensemble \cite{calabresealba}.
We characterize the clock's bias conditions, showing that in the absence of degeneracies, these depend solely on the quench parameter, though generally they also depend on the chosen observable. We demonstrate a stringent implication by linking the negativity of the response function of the chosen observable at the qubit's frequency to the clock bias condition.
In the examples provided, the presence of a dynamical phase transition and population inversion is decisive for powering the clock. This suggests that crossing the critical point of the phase transition is decisive for energy extraction from the battery. However, it remains unclear whether this is a universal condition in terms of the model and the choice of a particular class of observables, such as the order parameter.

\begin{acknowledgments}
The Authors would like to acknowledge fruitful discussions with  M. Mitchinson,  R. Silva and F. Meier. A.S. would like to acknowledge support from PNRR MUR project PE0000023- NQSTI and Quantera project SuperLink.
\end{acknowledgments}

\bibliography{TheBibliography}

\newpage
\widetext
\begin{center}
\textbf{\large Supplemental Materials: Powering a quantum clock with a non-equilibrium steady state}
\end{center}
\setcounter{equation}{0}
\setcounter{figure}{0}
\setcounter{table}{0}
\setcounter{page}{1}
\makeatletter
\renewcommand{\theequation}{S\arabic{equation}}
\renewcommand{\thefigure}{S\arabic{figure}}
\renewcommand{\bibnumfmt}[1]{[S#1]}
\renewcommand{\citenumfont}[1]{S#1}

\section{The dynamics of the clock}\label{dynamics}
In this section, we derive the ladder dynamics in the weak coupling regime, which implies $g\ll \epsilon_0$. \\
The following superoperator represents the effect of the battery on the qubit
\begin{equation}
    \mathcal{L}_\gamma=\gamma_\downarrow \mathcal{D}[\hat{\sigma}^-]+\gamma_\uparrow \mathcal{D}[\hat{\sigma}^+],
\end{equation}
where $\hat{\sigma}_\pm$ is the qubit raising and lowering operators, 
$\gamma_\pm$ are the transition rates of the qubit states due to the battery, and the dissipator is given by
\begin{equation}
    \mathcal{D}[\hat{A}]\hat{\rho}=\hat{A}\hat{\rho} \hat{A}^\dag-\frac{1}{2}\{\hat{A}^\dag \hat{A},\hat{\rho}\}.
\end{equation}
We describe the dynamics of the clock in the "no-click" subspace. Let $\hat{\rho}_0(t)$ be the density matrix conditioned on no spontaneous emission occurring up to time t. We assume the pointer is initially in the product state
\begin{equation}
    \hat{\rho}_0(0)=\hat{\rho}_Q(0)\otimes \ket{0}\bra{0}_w
\end{equation}
where $\hat{\rho}_Q(0)$ is the initial state of the qubit. The equation of motion describes the evolution of the conditional density operator:
\begin{equation}
    \frac{d\hat{\rho}_0}{dt}=i(\hat{\rho}_0 \hat{H}^\dag_{eff}-\hat{H}_{eff}\hat{\rho}_0)+\mathcal{L}_\gamma \hat{\rho}_0
\end{equation}
where the effective Hamiltonian is
\begin{equation}
    \hat{H}_{eff}=\hat{H}_0+\hat{H}_{int}+\hat{H}_{se}.
\end{equation}
Here, $\hat{H}_0=\hat{H}_Q+\hat{H}_L$ and $\hat{H}_{int}=\hat{H}_{QL}$ and the spontaneous emission term, which is non-Hermitian 
\begin{equation}
    \hat{H}_{eff}=-\frac{i\Gamma}{2}\hat{c}_{d-1}^\dag \hat{c}_{d-1}
\end{equation}
\subsection{The biased random walk model}
In the limit $\gamma\gg g,\Gamma$, we use the Nakajima-Zwanzig projection technique to derive an evolution equation for the conditional reduced density operator of the ladder
\begin{equation}
    \hat{\rho}_w(t)=Tr_Q[\hat{\rho}_0(t)]
\end{equation}
We introduce the projector into the separated state, defined as
\begin{equation}
    \mathcal{P}\hat{\rho}_0(t)=\hat{\rho}_w(t)\otimes \hat{\rho}_Q
\end{equation}
We rewrite the evolution equation of the conditional density matrix as
\begin{equation}
    \frac{d\hat{\rho}_0}{dt}=\mathcal{L}\hat{\rho}_0
\end{equation}
where
\begin{equation}
    \mathcal{L}=\mathcal{L}_0+\mathcal{H}_{se}+\mathcal{H}_{int}.
\end{equation}
The Hamiltonian superoperator is given by
\begin{equation}
    \mathcal{H}_{se}\hat{\rho}=i(\hat{\rho} \hat{H}^\dag_{se}-\hat{H}_{se}\hat{\rho}).
\end{equation}
We transform the density operator to a dissipative interaction picture defined by $\tilde{\hat{\rho}}(t)=e^{-\mathcal{L}_0 t}\hat{\rho}_0(t)$. The corresponding evolution of the superoperators is given by
\begin{equation}
    \tilde{\mathcal{H}}_{int}(t)=e^{-\mathcal{L}_0 t}\mathcal{H}_{int}
e^{\mathcal{L}_0 t}
\end{equation}
while the operator $\tilde{\mathcal{H}}_{se}$ stays the same, as it commutes with the operator $\mathcal{L}_0$. Starting from the perturbative argument in the coupling strength of the interaction g and $\Gamma$, we can write 
\begin{equation}
    \frac{d\mathcal{P}\tilde{\hat{\rho}}_0}{dt}=\tilde{\mathcal{H}}_{se}\mathcal{P}\tilde{\hat{\rho}}_0(t)+\int_0^t dt' \mathcal{P}\tilde{\mathcal{H}}_{int}(t)\tilde{\mathcal{H}}_{int}(t')\mathcal{P}\tilde{\hat{\rho}}_0(t')
\end{equation}
valid up to the second order. We then apply the Born-Markov approximation, assuming $\gamma_{\uparrow,\downarrow}\gg g,\Gamma$. The next steps involve expanding the commutator in the basis of the jump operators, whose eigenvalues are the energy differences between the levels of the system, tracing over the qubit states, and transforming back to the Schrödinger picture. Assuming no initial coherence, the resulting master equation for the population of the levels $p_w$ is
\begin{eqnarray}
    \frac{d p_w}{dt}=p_\downarrow \mathcal{D}[\hat{B}_w]p_w+p_\uparrow \mathcal{D}[\hat{B}_w^\dag]p_w\\ \nonumber
    -\frac{\Gamma}{2}(\ket{d-1}_w\bra{d-1}_w p_w+p_w \ket{d-1}_w\bra{d-1}_w)
\end{eqnarray}
in terms of the ladder jump operators $\hat{B}_w=\hat{c}^\dag_k \hat{c}_{k+1}$.
In our case the resulting expression for $p_\uparrow$ and $p_\downarrow$ is 
\begin{equation}
    p_\uparrow=2g^2\int_0^\infty dt e^{i\epsilon_w t}\langle \hat{\sigma}^-(t) \hat{\sigma}^+(0) \rangle
\end{equation}
and
\begin{equation}
  p_\downarrow=2g^2\int_0^\infty dt e^{-i\epsilon_w t}\langle \hat{\sigma}^+(t) \hat{\sigma}^-(0) \rangle  
\end{equation}
The ladder operators evolve according to the adjoint Liouvillian operator $\hat{\sigma}^-(t)=e^{\mathcal{L}^\dag_0 t}\hat{\sigma}^-(0)$, resulting in $\hat{\sigma}^-(t)=\exp[-i\epsilon_0-(\gamma_\uparrow+\gamma_\downarrow) t]\hat{\sigma}_-$. Inserting this expression into the definition of $p_{\uparrow,\downarrow}$ we have
\begin{equation}
    p_{\uparrow}=\frac{2g^2}{\gamma_\uparrow+\gamma_\downarrow}\langle \hat{\sigma}^- \hat{\sigma}^+ \rangle_s
\end{equation}
and
\begin{equation}
    p_{\downarrow}=\frac{2g^2}{\gamma_\uparrow+\gamma_\downarrow}\langle \hat{\sigma}^+ \hat{\sigma}^- \rangle_s
\end{equation}
where these averages are calculated in the stationary state of the qubit.\\
Now, considering the stationary state of the qubit:
using the representation of the qubit density matrix as $\hat{\rho}(t)=\frac{1}{2}\bigg(1+\langle \Vec{\hat{\sigma}}(t)\cdot\vec{\hat{\sigma}} \rangle\bigg)$ and inserting it into the Lindblad equation for the dynamics of the system qubit+battery only
\begin{equation}
    \frac{d\hat{\rho}_Q}{dt}=i[H_Q,\hat{\rho}_Q]+\mathcal{L}_\gamma \hat{\rho}_Q
\end{equation}
we obtain the following evolution equations for the averages of the spin operators
\begin{equation}
    \frac{d\langle \hat{\sigma}^+\rangle}{dt}=-(\gamma_\uparrow+\gamma_\downarrow) \langle \hat{\sigma}^+\rangle
\end{equation}
\begin{equation}
    \frac{d\langle \hat{\sigma}^-\rangle}{dt}=-(\gamma_\uparrow+\gamma_\downarrow) \langle \hat{\sigma}^-\rangle
\end{equation}
and for $\hat{\sigma}_z$
\begin{equation}
    \frac{d\langle \hat{\sigma}^z \rangle}{dt}=-(\gamma_\uparrow+\gamma_\downarrow) \langle \hat{\sigma}^z \rangle+\gamma_\uparrow-\gamma_\downarrow.
\end{equation}
As a consequence, the stationary state has no coherences and 
\begin{equation}
    \langle \hat{\sigma}_z \rangle_s=\frac{\gamma_\uparrow-\gamma_\downarrow}{\gamma_\uparrow+\gamma_\downarrow}
\end{equation}
and the diagonal matrix elements are
\begin{equation}
    \langle \hat{\sigma}^+ \hat{\sigma}^- \rangle_s=\frac{1}{2}\bigg(1+\frac{\gamma_\uparrow-\gamma_\downarrow}{\gamma_\uparrow+\gamma_\downarrow}\bigg)
\end{equation}
and
\begin{equation}
    \langle \hat{\sigma}^- \hat{\sigma}^+ \rangle_s=\frac{1}{2}\bigg(1-\frac{\gamma_\uparrow-\gamma_\downarrow}{\gamma_\uparrow+\gamma_\downarrow}\bigg).
\end{equation}
Inserting these expression in the definitions of $p_{\uparrow,\downarrow}$ yields
\begin{equation}
    p_\uparrow=\frac{g^2}{\gamma_\uparrow+\gamma_\downarrow}\bigg[1+\frac{\gamma_\uparrow-\gamma_\downarrow}{\gamma_\uparrow+\gamma_\downarrow}\bigg]
\end{equation}
and
\begin{equation}
    p_\downarrow=\frac{g^2}{\gamma_\uparrow+\gamma_\downarrow}\bigg[1-\frac{\gamma_\uparrow-\gamma_\downarrow}{\gamma_\uparrow+\gamma_\downarrow}\bigg].
\end{equation}
\\

\section{The calculation of the transition rates}\label{transitionrates}
\subsection{The XX spin chain}
Now, let us calculate the rate of the dissipative process, namely $\gamma_{\uparrow,\downarrow}$.
We write these rates in the following form \cite{breuer}
\begin{equation}\label{rates}
    \gamma_{\uparrow,\downarrow}=\frac{1}{4}\int_{-\infty}^\infty d\tau e^{\mp i \epsilon_0 \tau}Tr[\hat{\rho}_{st}\hat{J}(\tau/2)\hat{J}(-\tau/2)],
\end{equation}
as $\hat{\sigma}_x=\frac{\hat{\sigma}_+ +\hat{\sigma}_-}{2}$. After performing the quench in the parameter $V_i\rightarrow V_f$, leaving the field $\phi$ to zero, the correlator of the current is calculated in the stationary regime, which is constituted by the diagonal ensemble $\hat{\hat{\rho}}_{st}=\sum_{ki}p_k \hat{\pi}_n$, where the operator $\hat{\pi}_k$ is defined as $\hat{\pi}_k=\sum_i \ket{\psi_{ki}}\bra{\psi_{ki}}$ the projector on the k-th eigenstate of the final Hamiltonian and $p_k=\braketmatrix{\psi_0}{\hat{\pi}_k}{\psi_0}$, $\psi_0$ being the initial state.\\
As it is clearly shown in \cite{rossini}, this Hamiltonian can be written in the Fourier space by introducing free-fermionic operators and then diagonalized employing a Bogoliubov transform, which allows us to rewrite it as
\begin{equation}
    \hat{H}_B=\sum_{|k|<\pi/2}\epsilon_k(\phi)\hat{\vec{\Gamma}}^\dag_k\hat{\sigma}_z\hat{\vec{\Gamma}}_k
\end{equation}
where
\begin{equation}
    \epsilon_k(\phi)=\sqrt{[2t\cos(k-\phi)]+V^2}
\end{equation}
and 
\begin{equation}
    \hat{\vec{\Gamma}}_k=(\gamma_k^+, \gamma_k^-)
\end{equation}
the + and - sectors of the eigenvalues.\\
We write the correlator in the basis of the eigenstates of the post-quench Hamiltonian and we insert an identity operator
\begin{eqnarray}\label{correlator}
    Tr[\hat{\rho}_{st}\hat{J}(\tau/2)\hat{J}(-\tau/2)]=\sum_{ki}\sum_{k'i'} p_k^i \bra{ki}_f{\hat{J}(\tau/2)}\ket{k'i'}_f \\ \nonumber \bra{k'i'}_f{\hat{J}(-\tau/2)}\ket{ki}_f
\end{eqnarray}
where
\begin{equation}\label{current}
\hat{J}_{\phi=0} = -it \sum_{j=1}^L \left( \hat{b}^\dag_j \hat{b}_{j+1} - \text{h.c.} \right).
\end{equation}
Here the $p_k^i$ are defined as the overlap between the initial state and the eigenstates of the final Hamiltonian. If we take the initial state as being $\ket{\psi_0}:=(\gamma_k^+, \gamma_k^-)_i=(0,1)_i$ $\forall k$, i.e. the ground state, we can write this overlap as
\begin{equation}
    p_k^i=|\bra{ki}_f\ket{\psi_0}|^2
\end{equation}
This quantity can be evaluated using the Bogoliubov transform which relates the eigenstates of the post-quench and the initial Hamiltonian as
\begin{equation}
    \hat{\vec{\Gamma}}_k (V_f)=e^{i\Delta\Theta_k\hat{\sigma}^y}\hat{\vec{\Gamma}}_k (V_i)
\end{equation}
More conveniently, the complex exponential can be written as
\begin{equation}\label{Bogoliubovneq}
    \hat{\vec{\Gamma}}_k (V_f)=[\cos(\Delta\Theta_k )+i\hat{\sigma}^y\sin(\Delta\Theta_k )]\hat{\vec{\Gamma}}_k (V_i)
\end{equation}
with
\begin{equation}
    \Delta\theta_k=\Theta_k(V_f)-\Theta_k(V_i).
\end{equation}
Bearing in mind that the initial state is the ground state and satisfies $\langle{\hat{\gamma}^{+\dag}_{k,V_i}}\hat{\gamma}^{+}_{k,V_i}\rangle=\langle{\hat{\gamma}^{-\dag}_{k,V_i}}\hat{\gamma}^+_{k,V_i}\rangle=0$ and $\langle{\hat{\gamma}^{-\dag}_{k,V_i}}\hat{\gamma}^-_{k,V_i}\rangle=1$ and substituting Eq.(\ref{Bogoliubovneq}) implies
\begin{equation}
    p_k^+=\sin^2(\Delta\Theta_k),\hspace{1 cm} p_k^-=\cos^2(\Delta\Theta_k)
\end{equation}
The next step is to write the current operator (Eq. \ref{current}) in the $\ket{ki}_f$ basis. First of all, we employ a Jordan-Wigner transformation, mapping hard-core bosons into fermions 
\begin{equation}
\hat{b}^\dag_j=\exp\bigg(i\pi\sum_{l<j}\hat{a}^\dag_l\hat{a}_l\bigg)\hat{a}_j^\dag,
\end{equation}
under which the current reads
\begin{equation}
    \hat{J}=-it\sum_{j=1}^L \bigg(\hat{a}_j^\dag \hat{a}_{j+1}-\hat{a}_j^\dag \hat{a}_{j+1}\bigg)
\end{equation}
which is formally equivalent to the previous expression. Now we switch to momentum representation, by Fourier transforming the fermionic operators
\begin{equation}
    \hat{a}_j=\frac{1}{\sqrt{L}}\sum_k e^{-ikj}\hat{a}_k
\end{equation}
where the momentum variable takes values $k=\pm \pi(2n+1)/L$ and $n=0,...,L/2-1$ and the current takes form
\begin{equation}
    \hat{J}=-2t/L\sum_k \sin(k) \hat{a}_k^\dag \hat{a}_k.
\end{equation}
This expression admits a more compact notation
\begin{equation}
    \hat{J}=-2t/L\sum_k \sin(k) \hat{\vec{\Psi}}_k^\dag \hat{\sigma}_z \hat{\vec{\Psi}}_k.
\end{equation}
where we introduced the vector $\hat{\vec{\Psi}}_k=(\hat{a}_k, \hat{a}_{k+\pi})$, which separates the momentum into two sectors. Under the Bogoliubov transform diagonalizing the final Hamiltonian, which reads
\begin{equation}\label{Bogoliubov}
    \hat{\vec{\Gamma}}_k (V_f)=[\cos(\Theta_k )+i\hat{\sigma}^y\sin(\Theta_k )]\hat{\vec{\Psi}}_k
\end{equation}
where the angle of rotation $\Theta_k(V_f)$ is defined by the relation
\begin{equation}
    \tan[2\Theta_k(V_f)]=\frac{V_f}{2t\cos{k}},
\end{equation}
we have
\begin{equation}\label{representation}
    \hat{J}=\sum_k {\hat{\vec{\Gamma}}_k}^\dag \mathcal{J}_k \hat{\vec{\Gamma}}_k,
\end{equation}
where
\begin{equation}
\mathcal{J}_k=\frac{2t}{L}\sin{k}[\cos(2\Theta_k)\hat{\sigma}_z+\sin(2\Theta_k)\hat{\sigma}_x]
\end{equation}
Moreover, Eq. \ref{correlator} can be further simplified by expressing the temporal evolution of the current operator 
\begin{eqnarray}
    \sum_{ki}\sum_{k'i'} p_k^i \bra{ki}_f e^{-iH_{B,f}\tau/2}{\hat{J}(0)}e^{iH_{B,f}\tau/2}\ket{k'i'}_f \\ \nonumber \bra{k'i'}_f  e^{iH_{B,f}\tau/2}{\hat{J}(0)}e^{-iH_{B,f}\tau/2}\ket{ki}_f.
\end{eqnarray}
Expressing the current in the representation of Eq. \ref{representation}, one can write it in terms of the quench parameters as
\begin{eqnarray}
    \hat{J}=\sum_k \frac{4t^2}{L^2}\sin^2{k}\bigg[\cos^2(2\Theta_k)+\sin^2(2\Theta_k)\\ \nonumber(\cos^2(2\Delta\Theta_k) e^{-i2|\epsilon_k|\tau}+\sin^2(2\Delta\Theta_k)e^{i2|\epsilon_k|\tau})\bigg]
\end{eqnarray}
where 
\begin{equation}
    |\epsilon_k|=\sqrt{4t^2\cos^2{k}+V_f^2}
\end{equation}
We discretize this expression in the continuum limit for the momenta 
\begin{eqnarray}
    \hat{J}=\frac{2t^2}{\pi L}\int_{-\pi/2}^{\pi/2}dk \sin^2{k}\bigg[\cos^2(2\Theta_k)+\sin^2(2\Theta_k)\\ \nonumber (\cos^2(\Delta\Theta_k)e^{-i2|\epsilon_k|\tau}+\sin^2(\Delta\Theta_k)e^{i2|\epsilon_k|\tau})\bigg]
\end{eqnarray}
and plug it in the expression for the rates (Eq. \ref{rates}). We have three terms which appear. the first one reads
\begin{equation}
    (1)=\frac{2t^2}{L\pi}\int_{-\infty}^\infty ds e^{\mp i \epsilon_0 s}\int_{-\pi/2}^{\pi/2}dk \sin^2{k}\cos^2(2\Theta_k), 
\end{equation}
since $\cos(2\Theta_k)=\frac{2t\cos{k}}{|\epsilon_k(V_f)|}$
\begin{equation}
    (1)=\frac{1}{8 L}\bigg[V_f^2+2t^2-|V_f|\sqrt{V_f^2+4t^2}\bigg]\delta(\epsilon_0).
\end{equation}
From this expression, we can note that this term can be discarded, as it does not contribute for $\epsilon_0\neq 0$. The second term reads
\begin{eqnarray}
    (2)=\frac{2t^2}{\pi L}\int_{-\pi/2}^{\pi/2}dk \int_{-\infty}^\infty ds e^{ i (\mp\epsilon_0+2|\epsilon_k|) s}\sin^2{k}\\ \nonumber \sin^2(2\Theta_k)\sin^2(\Delta\Theta_k),
\end{eqnarray}
which means
\begin{eqnarray}
    (2)=\frac{2t^2}{\pi L}\int_{-\pi/2}^{\pi/2}dk \delta(2|\epsilon_k|\mp \epsilon_0)(1-\cos^2{k})\\ \nonumber \sin^2(2\Theta_k)\bigg(\frac{1-\cos(2\Delta\Theta_k)}{2}\bigg).
\end{eqnarray}
The argument of the delta function admits a zero only for $\gamma_\uparrow$. 
The third term is
\begin{eqnarray}
(3)=\frac{2t^2}{\pi L}\int_{-\pi/2}^{\pi/2}dk \int_{-\infty}^\infty ds e^{-i (\pm\epsilon_0+2|\epsilon_k|) s}\\ \nonumber \sin^2{k}\sin^2(2\Theta_k)\sin^2(\Delta\Theta_k).
\end{eqnarray}
Performing the first integral we obtain
\begin{eqnarray}
(3)=\frac{2t^2}{\pi L}\int_{-\pi/2}^{\pi/2}dk \delta(2|\epsilon_k|\mp \epsilon_0)(1-\cos^2{k})\sin^2(2\Theta_k)\\ \nonumber\bigg(\frac{1-\cos(2\Delta\Theta_k)}{2}\bigg).
\end{eqnarray}
\subsection{The Ising spin chain}
As it is known in the literature \cite{glen}, the model is exactly solvable in its fermionic formulation, after using the Jordan-Wigner transformations. Its expression is
\begin{equation}
    \hat{H}_B=-J\sum_{j=1}^L(\hat{c}_j\hat{c}^\dag_{j+1}+\kappa\hat{c}_j\hat{c}^\dag_{j+1}+h.c.)+h\sum_{j=1}^L(2\hat{n}_j-1),
\end{equation}
and $\hat{n}_j=\hat{c}^\dag_j \hat{c}_j$
where we assumed open boundary conditions. The Fourier transform is defined up to a phase, which does not modify the commutation relations, i.e.
\begin{equation}
    \hat{c}_k=\frac{e^{-i\phi}}{\sqrt{L}}\sum_{j=1}^L e^{-ikj}\hat{c}_j.
\end{equation}
Now the Hamiltonian reads
\begin{equation}
    \hat{H}_B=\sum_k \hat{H}_k,
\end{equation}
where
\begin{equation}    \hat{H}_k=\sum_{\alpha,\alpha'}\hat{\Psi}^\dag_{k,\alpha} (\boldsymbol{H}_k)_{\alpha\alpha'} \hat{\Psi}_{k,\alpha'}
\end{equation}
in terms of the vector
\begin{equation}
   \hat{\Psi}_k=
\begin{pmatrix}
\hat{c}_k\\
\hat{c}^\dag_{-k}
\end{pmatrix}
\end{equation}
The correspondent values of k are in the range
\begin{equation}
    k=\pm\frac{(2n+1)\pi}{L}
\end{equation}
The matrix $\boldsymbol{H}_k$ can be represented in terms of pseudo-Pauli matrices as 
\begin{eqnarray}
    \boldsymbol{H}_k=2(-\kappa \sin(2\phi)\sin(k)\hat{\tau}_x+\kappa \cos(2\phi)\sin(k)\hat{\tau}_y\\ \nonumber+(h-\cos(k))\hat{\tau}_z)
\end{eqnarray}
The function of the variable $\phi$ is to highlight the freedom we have in choosing the second axis of the spin in the xy plane.\\
This Hamiltonian is diagonalized using the Bogoliubov transform, having eigenvalues $\epsilon_{k\pm}=\pm |\epsilon_k|$, according to the parity sector, and
\begin{equation}
    |\epsilon_k|=2\sqrt{(h-\cos(k))^2+\kappa^2 \sin^2(k)}\geq 0
\end{equation}
In terms of the transformed eigenstates, now the Hamiltonian reads
\begin{equation}
\hat{H}_k=\epsilon_k(\hat{\gamma}_k^\dag \hat{\gamma}_k-\hat{\gamma}_{-k}^\dag \hat{\gamma}_{-k}-1)
\end{equation}
Now, we choose the coupling between the battery and the qubit as
\begin{equation}
    \hat{H}_{QB}=g_\sigma \hat{\sigma}^x \sum_{j=1}^L \hat{\sigma}_j^z,
\end{equation}
therefore the transition rates read
\begin{equation}
    \gamma_{\uparrow,\downarrow}=\frac{g_\sigma}{4}\sum_{j=1}^L \sum_{j'=1}^L \int_{-\infty}^\infty ds e^{\mp i \epsilon_0 s} Tr[\hat{\rho}_{st} \hat{\sigma}_z^j(s/2)\hat{\sigma}_z^{j'}(-s/2)]
\end{equation}
As before, this can be decomposed in the basis of the eigenvalues of the final parameter of the quench, which in this case is the transverse magnetic field h
\begin{eqnarray}
    \sum_{j=1}^L \sum_{j'=1}^L Tr[\hat{\rho}_{st} \hat{\sigma}^j_z(\tau/2)\hat{\sigma}^{j'}_z(-\tau/2)]=\\ \nonumber \sum_{ki}\sum_{k'i'}  \bra{ki}\hat{\sigma}_z(\tau/2)\ket{k'i'}\bra{k'i'}\hat{\sigma}_z(-\tau/2)\ket{ki}
\end{eqnarray}
The operator $\hat{\sigma}_z$ can be written in the fermionic operators as
\begin{equation}
    \hat{\Psi}^\dag_k 2\hat{\sigma}_z \hat{\Psi}_k.
\end{equation}
Upon performing the Bogoliubov transformation, the operator reads
\begin{equation}
    2\hat{\Gamma}^\dag_k (\cos(2\Theta_k)\hat{\sigma}_z+\sin(2\Theta_k)\hat{\sigma}_x) \hat{\Gamma}_k
\end{equation}
where the new basis vectors are defined by
\begin{equation}
    \hat{\Gamma}_k=U_k^\dag \hat{\Psi}_k
\end{equation}
and
\begin{equation}
   U_k^\dag=e^{i\hat{\sigma}_y \Theta_k} 
\end{equation} by the angle defined as
\begin{equation}
    \tan(2\Theta_k)=\frac{\kappa \sin(k)}{h-\cos(k)}
\end{equation}
First of all, the overlap $p_{k}^i$ between the initial state, the ground state of the initial Hamiltonian, is 
\begin{equation}
    p_k^+=\sin^2(\Delta\Theta_k)
\end{equation}
and
\begin{equation}
    p_k^-=\cos^2(\Delta\Theta_k).
\end{equation}
where $\Delta\Theta_k=\Theta_{k,f}-\Theta_{k,i}$.
Finally, we can substitute this relation into the correlator, which reads
\begin{eqnarray}
    \frac{4}{L^2}\sum_k\bigg[\cos^2(2\Theta_k)+\sin^2(2\Theta_k)(\cos^2(\Delta\Theta_k)e^{-2i|\epsilon_k|\tau}\\ \nonumber +\sin^2(\Delta\Theta_k)e^{+2i|\epsilon_k|\tau})\bigg].
\end{eqnarray}
We now take the continuum limit
\begin{eqnarray}
    \frac{2}{\pi L}\int_{-\frac{\pi}{2}}^{\frac{\pi}{2}} dk\bigg[\cos^2(2\Theta_k)+\sin^2(2\Theta_k)\\ \nonumber (\cos^2(\Delta\Theta_k)e^{-2i|\epsilon_k|\tau} +\sin^2(\Delta\Theta_k)e^{+2i|\epsilon_k|\tau})\bigg].
\end{eqnarray}
Let us now take into account each of these terms. Given that $\cos(2\Theta_k)=\frac{2(h_i-\cos k)}{\epsilon_k(h_i)}$, the first contribution to the rates reads
\begin{equation}
    (1)=\frac{2}{\pi L}\delta(\epsilon_0)\int_{-\pi/2}^{\pi/2} dk \frac{4(h_i-\cos k)^2}{\epsilon_k(h_i)^2},
\end{equation}
which is clearly equal to zero, as $\epsilon_0>0$. The second term reads
\begin{equation}
    (2)=\frac{2}{\pi L}\int_{-\pi/2}^{\pi/2} dk \delta(2|\epsilon_k|\mp\epsilon_0)\sin^2(2\Theta_k)\sin^2(\Delta\Theta_k),
\end{equation}
while the third
\begin{equation}
    (3)=\frac{2}{\pi L}\int_{-\pi/2}^{\pi/2} dk \delta(2|\epsilon_k|\pm\epsilon_0)\sin^2(2\Theta_k)\cos^2(\Delta\Theta_k)
\end{equation}
The argument of the delta function implies that (2) contributes only to $\gamma_\uparrow$, while (3) to $\gamma_\downarrow$.\\
First of all, $\sin(2\Theta_k)=\tan(2\Theta_k)\cos(2\Theta_k)=\frac{2\kappa \sin(k)}{\epsilon_k(h_f)}$ and we make use of the relation 
$\cos^2(\Delta\Theta_k)=\frac{1+\cos(2\Delta\Theta_k)}{2}$ and \begin{equation}\begin{split}
&\cos(2\Delta\Theta_k)=\cos(2\Theta_{k,f})\cos(2\Theta_{k,i})+\sin(2\Theta_{k,f})\sin(2\Theta_{k,i})=\\&\frac{4(h_i-\cos(k))(h_f-\cos(k))+(2\kappa \sin(k))^2}{|\epsilon_k(h_f)||\epsilon_k(h_i)|}
\end{split}\end{equation}
The delta function admits a solution for $|\epsilon_k|=\frac{\epsilon_0}{2}$, which implies
\begin{equation}
    (h_f-\cos{k})^2+\kappa^2 \sin^2(k)=\epsilon_0^2/16.
\end{equation}
This equation admits two solutions, parametrized as $u_\pm=\cos{k_{\pm}}$, i.e.
\begin{equation}\label{solutions}
    u_\pm=\frac{h_f\pm\sqrt{\kappa^2 h_f^2-(1-\kappa^2)(\kappa^2 -\epsilon_0^2/16)}}{(1-\kappa^2)}
\end{equation}
The conditions for the existence of this solution are detailed in the Main Article.
\section{The imaginary part of the response function}\label{Staggered}
In general, suppose we start from a Hamiltonian depending on the initial parameter of the quench $\lambda_i$, $\hat{H}(\lambda_i)$ and the system is prepared in the initial state $\ket{\psi_0}$. Then, the parameter is quenched to $\lambda_f$ and subject to an external time-dependent perturbation $\hat{V}(t)=h(t)\hat{B}$. For any generic observable $A$, the expectation value in the linear response regime reads
\begin{equation}
    \langle \hat{A} \rangle(t)=\langle \hat{A} \rangle(0)+\int_0^t dt' \chi_{AB}(t,t')h(t')
\end{equation}
where
\begin{equation}
    \chi_{AB}(t,t')=-i\theta(t-t')\bra{\psi_0}[\hat{A}(t),\hat{B}(t')]\ket{\psi_0}.
\end{equation}
We rewrite the latter in terms of the time variables $T=\frac{t+t'}{2}$ and $\tau=t-t'$ and average over the centre-of-mass time variable
\begin{eqnarray}
    \bar{\chi}_{A,B}(\tau)=-i\theta(\tau)\lim_{T\rightarrow\infty}\frac{1}{T}\int_0^T dt' Tr\bigg\{\hat{\rho}(0)\\ \nonumber \bigg[\hat{A}(T+\tau/2),\hat{B}(T-\tau/2)\bigg]\bigg\}
\end{eqnarray}
where the initial density matrix is $\hat{\rho}(0)=\ket{\psi_0}\bra{\psi_0}$. The expression inside the parenthesis can be rewritten as $e^{-i\hat{H}_f T}[\hat{A}(\tau/2),\hat{B}(-\tau/2)]e^{i\hat{H}_f T}$. After averaging, the trace over $\hat{\rho}(0)$ can be performed under the diagonal ensemble, representing the stationary state $\hat{\rho}_{st}=\sum_{n,r}p_n\ket{\psi_{n,r}}\bra{\psi_{n,r}}$, where $p_n$ is the overlap with the initial state and r a degeneracy index. At the end, one obtains
\begin{equation}
    \bar{\chi}_{A,B}(\tau)=-i\theta(\tau)\sum_{n}p_n \bra{\psi_n}[\hat{A}(\tau/2),\hat{B}(-\tau/2)]\ket{\psi_n}.
\end{equation}
In our case, $\bar{\chi}(\omega)$ represents the Fourier transform of $\bar{\chi}_{A,B}(\tau)$ and $\bar{\chi}''(\omega)=-Im\bar{\chi}(\omega)$.\\
\section{The estimation of battery's lifetime}
The total number of ticks over the battery lifespan is given by the energy available in the battery divided by the energy of the photon emitted $E_{ph}=(d-1)\epsilon_0$
\begin{equation}
    N_p=\frac{E_{av}}{E_{ph}}
\end{equation}
Therefore the total time is
\begin{equation}
    T^*=\frac{E_{av}\tau}{E_{ph}}
\end{equation}
The energy available in the battery is 
\begin{equation}
    E_{av}=\sum_k 2\epsilon_k \delta(2\epsilon_k-\epsilon_0)\cos^2(\Delta\Theta_k)
\end{equation}
In  the continuum limit, this quantity reads
\begin{equation}
    E_{av}=\frac{L}{2\pi}\int dk 2\epsilon_k \delta(2\epsilon_k-\epsilon_0)\cos^2(\Delta\Theta_k)
\end{equation}
We pass to the variable $\epsilon_k$, introducing an integration over a spherical shell of energy $\epsilon_k\in\Omega=[\epsilon_0/2,\epsilon_0/2+\Delta]$, with an infinitesimal thickness $\Delta$
\begin{eqnarray}
E_{av}=\frac{L}{2\pi}\int d\epsilon_k \nu(\epsilon_k) 2\epsilon_k \cos^2(\Delta\Theta_k)\\ 
\simeq\frac{L}{2\pi}\epsilon_0 \nu(\epsilon_0/2) \cos^2(\Delta\Theta_k)|_{k=k^*}\Delta
\end{eqnarray}
Normalizing by $\Delta$ one has
\begin{equation}
    E_{av}=\frac{L}{2\pi}\epsilon_0 \nu(\epsilon_0/2) \cos^2(\Delta\Theta_k)|_{k=k*}
\end{equation}
Therefore, the total lifespan of the battery reads
\begin{equation}
T^* = -\frac{\gamma_\uparrow+\gamma_\downarrow}{\bar{\chi}''(\epsilon_0)} \frac{L}{2\pi} \rho(\epsilon_0/2) \cos^2(\Delta\Theta_k) \bigg|_{k=k^*},
\end{equation}
\end{document}